# Effect of Electron-Phonon Interactions on Three-Level QD-based Spaser: Linear and Quadratic Potentials

Ankit purohit[*, £], Vishvendra Singh Poonia[†, ††, †††], Akhilesh Kumar Mishra [£, †††, **]

[£]*Department of Physics, Indian Institute of Technology Roorkee, Roorkee-247667, Uttarakhand, India*
[††]*Department of Electronics and Communication Engineering, Indian Institute of Technology Roorkee, Roorkee-247667, Uttarakhand, India*
[†††]*Centre for Photonics and Quantum Communication Technology, Indian Institute of Technology Roorkee, Roorkee- 247667, Uttarakhand, India*

[*]a_purohit@ph.iitr.ac.in, [†]vishvendra@ece.iitr.ac.in, [**]akhilesh.mishra@ph.iitr.ac.in

**Abstract:** In this article, a spaser (surface plasmon amplification by stimulated emission of radiation) system consisting of a metal nanoparticle surrounded by a large number of quantum dots (QDs) is studied. Usually, the effect of electron-phonon interaction is neglected in the spaser related literature. But some gain media, attributed by the large Raman scattering cross-section, exhibit stronger electron-phonon interaction. No such study has been performed for a QD-based spaser. Hence, it warrants investigation of the same in spaser system. In the present work, we investigate the effects of electron-phonon interaction on a three-level QD-based spaser. We consider two types of interaction potentials, linear and quadratic, and analyze their effects individually. First, we focus on the linear electron-phonon interaction that perturbs the electrons present in the excited state. This yields a periodic steady-state number of localized surface plasmon (LSP). The accompanying analytic solution reveals that the population inversion of the gain medium depends on the linear potential strength (Frohlich constant) but does not affect the threshold of spaser considerably for the given numerical parameters. In addition to the LSP, phonons may be generated during this process, the temporal dynamics of which are also detailed. Initially, the number of phonons exhibit decaying periodic oscillations, whose amplitude depends on the strength of the electron-phonon interaction. Under continuous pumping, at later times, the number of phonons reaches a steady-state value, which may find application in realization of continuous phonon nanolasers. Further, the effect of the quadratic potential is studied phenomenologically by increasing the excited-state decay rate. This results in a large number of LSP and an intense spaser spectrum.

## I. Introduction

QDs (or dye molecules) are the typical gain media that are used in the spaser experiments [1-3]. Usually, the electron-phonon interactions are ignored while studying spaser, but this interaction could be profound in gain media having large Raman gain coefficients [4, 5]. The strength of this interaction depends on the intrinsic properties of sample (gain medium), excitation conditions and the environmental factor (i.e., temperature or pressure) [6]. It is worth noting that the effect of electron-phonon interaction has been demonstrated experimentally on the observation of emission-line broadening of photoluminescence spectra, sideband linewidth broadening in resonance fluorescence and phonon induced damping of Rabi oscillations and so on [ 7-9]. Importantly, strong electron-phonon interactions are the cause of the enhancement in Raman signal, photosynthetic light harvesting among others [5, 10].

Plasmon resonance has been studied extensively in literature. Over the last two decades, quantum nanoplasmonics has been the topic of immense research interest [11]. Plasmon resonances find applications in different fields of applied sciences, including controlling chemical reactions, improving bio-sensing, increasing nonlinear effects and efficient solar energy conversion [11-16]. Plasmonic resonances can be excited in forms of surface plasmon polariton (SPP) and Localized plasmon resonance (LSP). Under the presence of external field, SPP excites at bulk metal-dielectric interface while LSP is supported in metal nanoparticles [1]. Both structures have been used in realizing plasmon lasing. LSP holds applications in devices confined from all three dimensions. At present, the size mismatch between the electronic and photonic devices has become a significant barrier in realizing on-chip electro-optic devices [16]. Conventional laser resonator dimensions are limited by the diffraction limit, which allows for the resonator's minimum size as λ/2 to satisfy the resonance condition, where λ is the wavelength of the light in the medium. The emerging field of quantum nanoplasmonics, in which light can focus on subwavelength regimes, offers hope for realizing even smaller nano-devices with a possibility of on chip integration [3, 11-17].

D. Bergman and M.I. Stockman were the first to propose spaser in 2003 [17,18]. In the spaser, the gain medium around the metal nanoparticle is pumped during operation, and the excitation energy is transferred to a nearby resonant plasmonic mode. Surface plasmons already present in the metal nanoparticles are amplified and this drives the transition of the gain medium through the feedback loop, resulting in the generation of a large number of coherent LSPs. In 2010, Stockman presented semiclassical theory of spasers in which the gain medium is considered as a two-level system [18]. The results of this theory qualitatively match with the experiments [1]. In a two-level system, the continuous feedback from the LSP mode causes gain saturation. Hence, the number of LSP in the spaser remains very low. Moreover, spaser has very high threshold rate, which limits its application. To overcome these problems, Dorfman et al. suggested that an additional external coherent light source in a three-level gain medium could increase the number of LSP and reduce the threshold [19, 20]. While implementing this it was taken care of that the external coherent light source spectrum should not interfere with the plasmon spectrum. But coherence is a very fragile phenomenon; therefore, it is difficult to drive the transition coherently using external light sources. In 2018, Song et al., however, without using additional external coherent light, experimentally demonstrated the operation of a spaser using a three-level gain medium [2]. The long lifetime of a three-level system increases the probability of a population inversion. In addition, because of the spin-forbidden transition from the singlet state to the triplet state, loss caused by stimulated absorption in the gain medium was almost eliminated. These two features in the three-level system reduces the pump threshold compared with the two-level spaser system [2].

Several effects have been ignored in the theoretical analysis of the spaser to minimize numerical complexity [19]. One of these is the negligence of the electron-phonon interaction [4-6]. In addition to the spontaneous emission rate, the decay rate of QDs depends on many scattering processes, such as electron-electron scattering, acoustic and optical phonon scattering, scattering by ionized impurities, and scattering from interface roughness [21]. In literature, the contribution of electron-phonon interaction is accounted by considering the dephasing rate for phonon dissipation phenomenologically, which results in the suppression of non-diagonal elements of density matrix. Therefore, the spaser efficiency is reduced [22, 23]. Recently, Tereshchenkov et. al. studied the electron-phonon interaction effects on the laser mode of a two-level QD gain medium [4]. They showed that different operational regimes of lasing are possible with the cavity frequency and electron-phonon interaction strength (Frohlich constant).

In this article, we consider a microscopic model to explain the electron-phonon interaction in QD-based spaser [4, 23]. Herein, we discuss two types of electron-phonon interactions, linear and quadratic. The linear electron-potential strength is of the order of $\approx 10^{-2} ev$, while the strength of quadratic potential is of the order of $\approx 10^{-3} ev$ [4, 24-26]. In linear potential, with the coherent LSP, coherent phonons may be generated, which is useful in coherent phonon nanolasers. In our analysis, we fix the plasmon frequency and study the spaser dynamics in the presence of Frohlich constant and pump rate. On the other hand, it is very cumbersome to study the quadratic potential theoretically [24-26]. For simplicity, the effect of quadratic potential is included phenomenologically by increasing the upper-level decay rate [23]. The quadratic coupling of phonons works as a thermal bath that increases the transition of higher exciton levels [27]. This results in a large number of LSP and an intense spaser spectrum. We believe that this study unveils further the role of phonons in spaser system.

The article has been arranged as follows: In sec. II, theoretical and mathematical models are described. We use the Frohlich Hamiltonian to describe the linear electron-phonon interaction; however, the contribution of the quadratic phonon-electron interaction is added phenomenologically. In sec. III, the numerical results are presented, wherein spaser and phonons dynamics are discussed. Sec. IV summarizes the work.

## II. Mathematical and theoretical model

We have considered a 40 nm silver nanoparticle surrounded by a large number of quantum dots, as shown schematically in fig. 1(a) [19, 20]. The QD is assumed to mimic a three-level gain medium, where an incoherent pump source couples $|1\rangle \rightarrow |3\rangle$ levels of the QD, as depicted in fig. 1 (b). In this study the QDs are assumed to have large Raman scattering cross-sections where electron-phonon interactions are significant [4]. Optical phonons interact with the electrons present in level $|2\rangle$ [4, 23]. The transition energy of $|2\rangle \rightarrow |1\rangle$ level is directly transferred to the nearby LSP mode with strength $\tilde{\Omega}_b$.

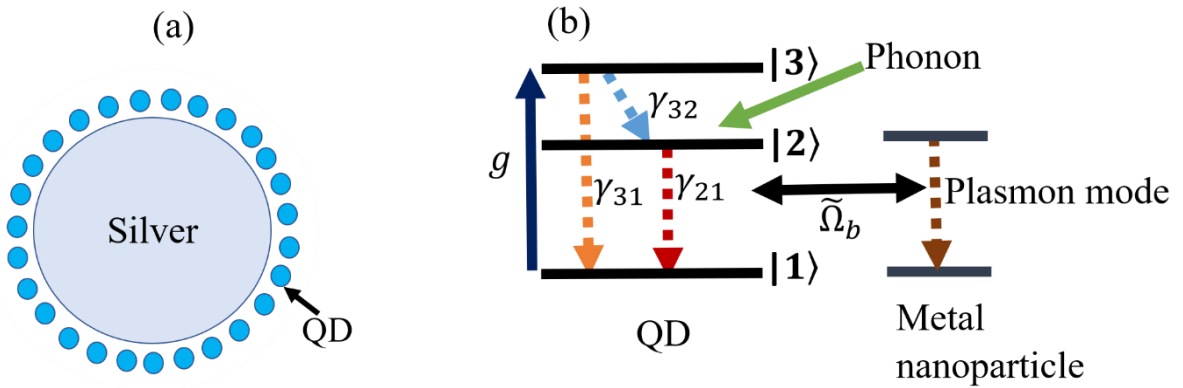

**Fig. 1: (a).** Schematic of a silver nanoparticle surrounded by quantum dots, **(b).** The incoherent light source couples $|1\rangle \rightarrow |3\rangle$ transition with rate g, while $|2\rangle \rightarrow |1\rangle$ transition couple the nearby dipolar plasmon mode.

In the present work, we have treated the quantum dots quantum mechanically while the plasmon mode is dealt semi-classically. Hence, the plasmon mode is considered through creation and annihilation operators $\hat{a}_n^\dagger$ and $\hat{a}_n$ respectively. The approximation used the theoretical development is that all operators behave as if they were classical numbers, which is valid as long as the number of quantum dot is large compared to one [23]. Hence, the LSP annihilation operator $\hat{a}_n$ is written as $\hat{a}_n = a_{0n}e^{-i\omega t}$, where $a_{0n}$ is slowly varying amplitude of LSP. The number of coherent LSPs per spasing mode is given by $N_n = |a_{0n}|^2$.

The system Hamiltonian in rotating wave approximation is written as-

$$H = H_n + H_b + H_g + H_{ng} + H_{bg}, \tag{1}$$

where $H_n$ is the Hamiltonian for the LSP mode, $H_b$ is the Hamiltonian for the phonon in the harmonic approximation [4], $H_g$ is the Hamiltonian for the quantum dot, $H_{ng}$ denotes the interaction Hamiltonian for the plasmon and quantum dots interaction. The last term in above Hamiltonian, $H_{bg}$ represents the interaction Hamiltonian for electron-phonon interaction and is given by $H_{bg} = |2\rangle\langle 2|(V_l + V_q)$. The linear potential term $V_l$ is responsible for phonon absorption/emission. However, the quadratic potential $V_q$ supports the transition to the higher excitonic level induce by the scattering of thermal phonons [23, 27]. Under the influence of external field, the charge distribution in QD is rearranged. As a result, excitons are formed and the lattice of the host material give response as $H_{bg}$. In the following, we will separately consider the effects of linear and quadratic potential terms on spaser mode.

*Electron-phonon linear interaction potential*: The linear potential is described as [23]

$$V_l = \hbar\Omega_g(\hat{b} + \hat{b}^\dagger), \tag{2}$$

where $\Omega_g$ is defined as linear electron-phonon interaction strength (Frohlich constant). The operators $\hat{b}$ and $\hat{b}^\dagger$ are the annihilation and creation operators for phonon. The phonon annihilation operator can be written as $\hat{b} = be^{-i\omega_m t}$, where $b$ is slowly varying amplitude. The linear phonon coupling describes real phonon processes that can be interpreted as an exciton-induced displacement of the host lattice. Eq. (1) takes the following form in presence of linear electron-phonon interaction:

$$H = \hbar\omega_n \hat{a}_n^\dagger \hat{a}_n + \sum_p \left[ \hbar\omega_m^p \hat{b}^\dagger \hat{b} + \sum_{i=1,2,3} \hbar\omega_i^p |i\rangle\langle i| + \hbar\Omega_g^p (\hat{b} + \hat{b}^\dagger)|2\rangle\langle 2| \right. $$
$$\left. + \left(\hbar\tilde{\Omega}_b^p |2\rangle\langle 1| + c.c\right) \right], \tag{3}$$

where $\Omega_b^p = -A_n d_{21}^p \nabla\varphi_n a_{0n}/\hbar = \tilde{\Omega}_b^p a_{0n}$ is the Rabi frequency for the spasing transition $|2\rangle \to |1\rangle$, and $\Omega_g^p$ is the strength of phonon interaction with $|2\rangle$ level electrons. Here, $p$ is the index of quantum emitter (gain medium), and $d^{(p)}$ is the transitional dipole moment. We will omit the index $p$ in forth coming discussion because all QDs are assumed to be identical in our theoretical analysis.

The emitter density matrix element for $p^{th}$ chromophore satisfies the Liouville—von Neumann equation,

$$\dot{\hat{\rho}} = -\frac{i}{\hbar}[H, \hat{\rho}] + \mathcal{L}(\hat{\rho}), \tag{4}$$

where $\mathcal{L}$ is the Lindblad operator, which quantifies the dissipation of QDs and incoherent pumping part. The Lindblad operator is expressed as

$$\mathcal{L}(\hat{\rho}) = \frac{\gamma_{32}}{2}(2|2\rangle\langle 3|\hat{\rho}|3\rangle\langle 2| - |2\rangle\langle 2|\hat{\rho} - \hat{\rho}|2\rangle\langle 2|) + \frac{\gamma_{21}}{2}(2|1\rangle\langle 2|\hat{\rho}|2\rangle\langle 1| - |1\rangle\langle 1|\hat{\rho} - \hat{\rho}|1\rangle\langle 1|)$$
$$+ \frac{\gamma_{31}}{2}(2|1\rangle\langle 3|\hat{\rho}|3\rangle\langle 1| - |1\rangle\langle 1|\hat{\rho} - \hat{\rho}|1\rangle\langle 1|)$$
$$+ \frac{g}{2}(2|3\rangle\langle 1|\hat{\rho}|1\rangle\langle 3| - |3\rangle\langle 3|\hat{\rho} - \hat{\rho}|3\rangle\langle 3|). \tag{5}$$

To make the Hamiltonian time independent, following transformations have been implemented- $\hat{\rho}_{31} = \rho_{31}e^{-i\omega_{31}t}$; $\hat{\rho}_{21} = \rho_{21}e^{-i\omega t}$ and $\hat{a}_n = a_{0n}e^{-i\omega t}$, where $\rho_{31}$, $\rho_{21}$, and $a_{0n}$ are the slowly varying amplitudes.

The density matrix elements for any $p^{th}$ chromophore have following elements-

$$\dot{\rho}_{11} = \gamma_{21}\rho_{22} + \gamma_{31}\rho_{33} - g\rho_{11} - i[\Omega_b^*\rho_{21} - \Omega_b\rho_{21}^*], \tag{6A}$$

$$\dot{\rho}_{33} = -(\gamma_{31} + \gamma_{32})\rho_{33} + g\rho_{11}, \tag{6B}$$

$$\dot{\rho}_{21} = -\Gamma_{21}\rho_{21} + i\Omega_b[\rho_{22} - \rho_{11}] - i\Omega_g[\hat{b}^\dagger + \hat{b}]\rho_{21}, \tag{6C}$$

$$\dot{\rho}_{32} = -\Gamma_{32}\rho_{32} + i\Omega_b^*\rho_{31} + i\Omega_g[\hat{b}^\dagger + \hat{b}]\rho_{32}, \tag{6D}$$

$$\dot{\rho}_{31} = -\Gamma_{31}\rho_{31} + i\Omega_b\rho_{32}. \tag{6E}$$

These density matrix elements also satisfy the conservation rule,

$$\rho_{11} + \rho_{22} + \rho_{33} = 1. \tag{7}$$

Since the LSP has dipolar mode, the Heisenberg equation of motion for LSP has similar form as that for two-level gain medium,

$$\dot{a}_{0n} = -(\gamma_n + i\Delta_n)a_{0n} - i\sum_p \rho_{21}^p \tilde{\Omega}_b^p, \tag{8}$$

where $\gamma_n$ is LSP relaxation rate, $\Delta_n = \omega_{21} - \omega_n$, and $\tilde{\Omega}_b^p = \Omega_b^p/a_{0n}$ is single plasmon Rabi frequency. The phonon operator has following form-

$$\dot{\hat{b}} = -(\gamma_b + i\omega_b)b - i\Omega_g \rho_{22}, \tag{9}$$

where $\Gamma_b = \gamma_b + i\omega_b$, $\gamma_b$ is phonon decay rate and $\omega_b$ is phonon frequency. The steady-state $\left(\dot{\rho}_{ij} = 0,\ \dot{a}_{on} = 0,\ \dot{\hat{b}} = 0\right)$ solution of eq. (6A) - (6E), (7), (8) and (9) gives,

$$\rho_{22} = \frac{\left[1 + \frac{\Gamma_n\Gamma_{21}}{N_p\tilde{\Omega}_b^2}\left(1 + \frac{g}{\gamma_{31} + \gamma_{32}}\right)\right]}{\left[1 + \left(1 + \frac{g}{\gamma_{31} + \gamma_{32}}\right)\left(1 - \frac{2\Gamma_n\Omega_g^2}{N_p\tilde{\Omega}_b^2\Gamma_b}\right)\right]}, \tag{10}$$

$$\rho_{11} = \frac{(1 - \rho_{22})}{\left(1 + \frac{g}{\gamma_{31} + \gamma_{32}}\right)}, \tag{11}$$

$$N_n = \frac{N_p}{2\Gamma_n}[-\gamma_{21}\rho_{22} - \gamma_{31}(1 - \rho_{11} - \rho_{22}) + g\rho_{11}]. \tag{12}$$

***Electron-phonon quadratic interaction potential:*** The quadratic potential is defined as

$$V_q = \hbar\kappa_g(\hat{b} + \hat{b}^\dagger)(\hat{b}' + \hat{b}'^\dagger). \tag{13}$$

The Hamiltonian for the quadratic potential is given as-

$$\mathrm{H} = \hbar\omega_n \hat{a}_n^\dagger \hat{a}_n + \sum_p \left[ \sum_{i=1,2,3} \hbar\omega_i^p |i\rangle\langle i| + \left(\hbar\tilde{\Omega}_b^p |2\rangle\langle 1| + c.c\right) \right], \tag{14}$$

where the contribution of quadratic phonon is added phenomenologically by increasing $\gamma_{32}$ in eq. 5. We would like to note that the quadratic electron-phonon interaction-is responsible for the dissipation and dephasing rates that depend on the excitation condition [5]. Since the dephasing rate reduces the efficiency of spaser, we, in our analysis, neglect the contribution of the dephasing rate [22]. If we somehow could manage the dephasing rate (temperature dependent), then the only contribution left is the population decay rate, which enhances the number of LSP in spaser systems.

## III. Results and Discussion

This section presents a detailed investigation of the effects of the phononic environment on input-output or spaser curve. The spaser dynamics above the threshold have been studied. We separately consider the effect of linear and quadratic electron-phonon interaction potentials on spaser. Unless otherwise specified, the following parameters have been used in our simulation: $\gamma_{21} = 4 \times 10^{12} s^{-1}$, $\gamma_{32} = 4 \times 10^{11} s^{-1}$, $\gamma_{31} = 4 \times 10^{10} s^{-1}$ [15,16]. The total number of quantum dots is $6 \times 10^4$ [19, 20]. For metal nanoparticle, we have used the following parameters- $\gamma_n = 5.3 \times 10^{14} s^{-1}$, $\Delta_n = 3 \times 10^{12} s^{-1}$, $\omega_n = 2.5\ eV$, and $\epsilon_d = 2.25$ [15,16]. The parameters for phonon are - $\omega_b = 0.025\ eV$, $\gamma_b = 3 \times 10^{11} s^{-1}$ and the Frohlich constant ($\Omega_g$) is of the order of $10^{-2}\ eV$ [4].

**(a). *Electron-phonon linear interaction*:** Here, we present the numerical simulation results obtained by solving the equations (6A) - (6E), 7, 8 and 9 in a steady-state ($\dot{\rho}_{ij} = 0$, $\dot{a}_{0n} = 0$, $\dot{\hat{b}} = 0$) regime. The pump rate at which population inversion ($\rho_{22} - \rho_{11} > 0$) occurs and system starts to spase (lase) is called threshold rate. Equations (10) and (11) show that the threshold rate depends on the Frohlich constant ($\Omega_g$); however, for the numerical parameters mentioned above, $N_n$ remain almost independent of $\Omega_g$ as depicted in Fig. 2 and so does population inversion. Fig. 2 shows that, as g increases, the steady-state values of $N_n$ increases; however, these values remain almost independent of $\Omega_g$. As mentioned previously, phonons only perturb the electrons in level $|2\rangle$, but the population inversion almost remains unaffected. We would like to note that for high Q cavity resonator, the linear potential may significantly vary the population inversion and therefore alter the lasing conditions [4].

To gain insight into the spaser dynamics, the time evolution of $N_n$ must be studied. Fig. 3 (a) and (b) show the temporal evolution of $\rho_{22}$ and $N_n$ at g =8THz pump rate for $\Omega_g = 0$ and $12 \times 10^{12} s^{-1}$. Population in level $|2\rangle$ is perturbed by the phonon and a periodic time harmonic steady state is observed for $\Omega_g = 12 \times 10^{12} s^{-1}$ as shown in Fig. 3(a). As a result, a periodic time harmonic steady state solution of $N_n$ is observed for the spaser, as displayed in Fig. 3(b). The amplitude of this oscillation depends on the $\Omega_g$ and the time period depends on $\omega_b$. The effects of these oscillations are profound on the spaser field. The spaser field is determined by taking Fourier transform of $a_{0n}(t)$ [4, 17], which is expressed as,

$$E(\omega) = \int_0^T a_{0n}(t) \, exp(i\omega_0 t) dt. \tag{15}$$

Fig. 3 (c) displays the variation of spaser field spectral envelope with $\Omega_g$. Initially, at $\Omega_g = 0$, when there is non-oscillatory steady-state solution in time, only a sharp resonance peak is observed in spaser field envelope. As $\Omega_g$ increases, the wings of the envelope broadens, and a small amplitude peak is observed at the spectral edges of the field envelope as shown in Fig. 3 (c). These small amplitude peaks owe it origin in periodic steady-state oscillation in time.

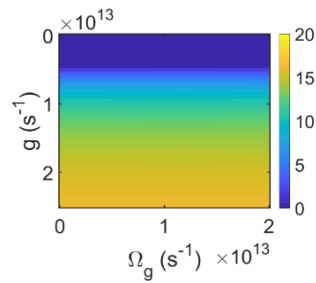

Fig. (2). Number of LSPs ($N_n$) for different values of g and $\Omega_g$ under steady state condition.

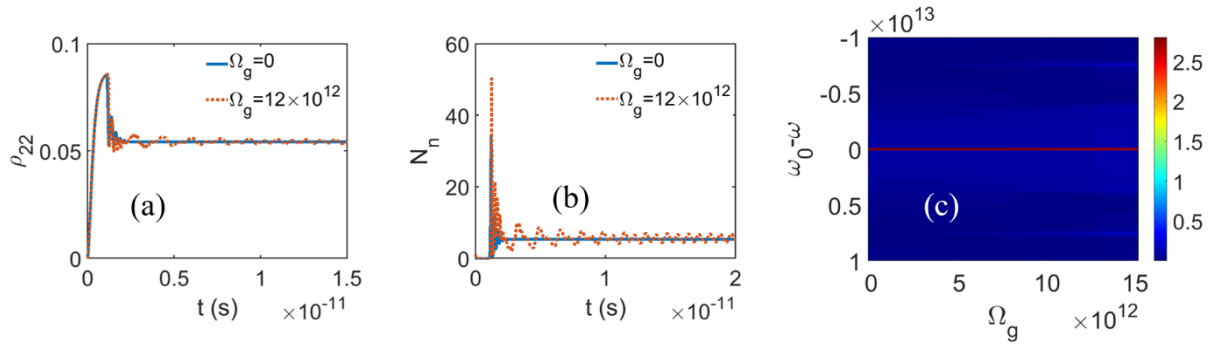

Fig. (3). Time evolution of (a) $\rho_{22}$, and (b) $N_n$ for $\Omega_g = 0$ and $12 \times 10^{12} s^{-1}$. (c) Evolution of spaser field envelope ($|E(\omega)|$) with $\Omega_g$ at g =8THz pump rate.

Besides the coherent LSPs, the steady-state phonons may also be generated during this process. Fig. 4 (a) shows the time evolution of $|b|^2$ for different values of $\Omega_g$ at a continuous pump rate of 8 THz. At t=0, there is no phonons (b=0). The amplitude of phonon oscillation, at later times, depends on the $\Omega_g$. Note that the time period of oscillation depends on the $\omega_b$ and does not depend on the $\Omega_g$. After certain time, the oscillations reach a steady state. As the $\Omega_g$ increases, large amplitude oscillation starts that lasts longer and ultimately relaxes to a large number of steady phonons. If there are some phonons at the initial time (t=0), then a large number of coherent phonons may be achieved, which can be used in phonon nanolasers. The phonon field is determined by taking Fourier transform of $b(t)$,

$$b(\omega) = \int_0^T b(t)\, exp(i\omega_0 t)dt. \tag{16}$$

Fig. 4 (b) shows the phonon field envelope for the different $\Omega_g$ values at a continuous pump rate of 8 THz. We obtained two peaks with different amplitudes in the phonon field envelope. The low amplitude peak appears at the phonon frequency $\omega_b$ while the large amplitude peak is observed on the blue side of spectrum, which is induced by the interaction between phonons and electrons present in level $|2\rangle$. The magnitudes of the peaks depend on the strength of $\Omega_g$. When we fix the $\Omega_g$ and increases the $\omega_b$, the separation between low and high amplitude peaks increases.

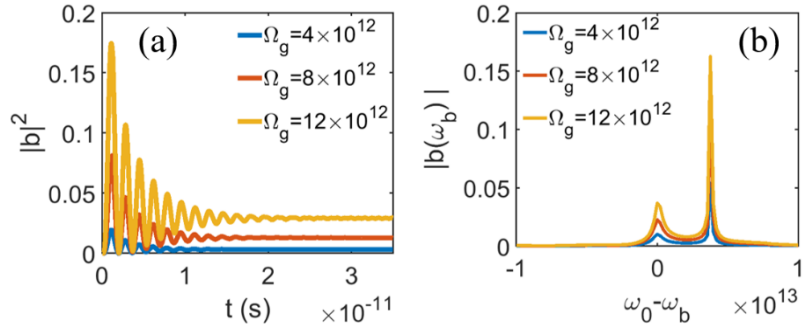

Fig. (4). (a) Time evolution of $|b|^2$ (b) and phonon field envelope($|b(\omega_b)|$) for different $\Omega_g$ values at g = 8 THz continuous pump rate.

**(b). *Electron-phonon quadratic interaction*:** As mentioned above, the contribution of the electron-phonon quadratic interaction is included phenomenologically and is made stronger by choosing larger $\gamma_{32}$ rate. When the electrons reach the excited state, they obviously decay spontaneously, which is induced by vacuum quantum fluctuations. For spontaneous decay, we take $\gamma_{32} = 4 \times 10^{11} s^{-1}$ in our simulation [19, 20]. Next, we choose $\gamma_{32} = 6 \times 10^{11} s^{-1}$ and $8 \times 10^{11} s^{-1}$, that includes the effect of quadratic electron-phonon interaction along with the spontaneous decay. The quadratic electron-phonon interaction provides us an extra degree of freedom to engineer the efficient spaser system, as discussed below. Fig 5(a) shows that with pumping rate g, $N_n$ increases as $\gamma_{32}$ increases. Note that the threshold rate remains independent of variation of $\gamma_{32}$ as seen in Fig 5 (a). This is because larger $\gamma_{32}$ increases the population inversion which results in the generation of large number of LSP ($N_n$) in steady state. Fig. 5(b) shows the temporal evolution of $N_n$ when pumped continuously with 8 *THz* rate for different $\gamma_{32}$ values. Initially, $N_n$ exhibits oscillations which relaxes to steady state after certain time. For larger $\gamma_{32}$, the oscillations start earlier, and their amplitudes are also large. Moreover, larger amplitude oscillations take longer time to relax to steady state and the corresponding steady state values are also larger. These results show that the dynamics of spaser sensitively depends on $\gamma_{32}$.

Since there are large number of $N_n$ for large $\gamma_{32}$ values, we observe intense spaser spectrum for larger $\gamma_{32}$ values, as displayed in Fig. 5(c). The resonance frequency of spaser is expressed as [12,15]-

$$\omega = \frac{\omega_{21}\gamma_n + \omega_n\gamma_{21}}{\gamma_{21} + \gamma_n}, \tag{17}$$

which depends on the decay rates ($\gamma_n$ and $\gamma_{21}$) and resonance frequencies ($\omega_n$ and $\omega_{21}$) of metal nanoparticle and QDs.

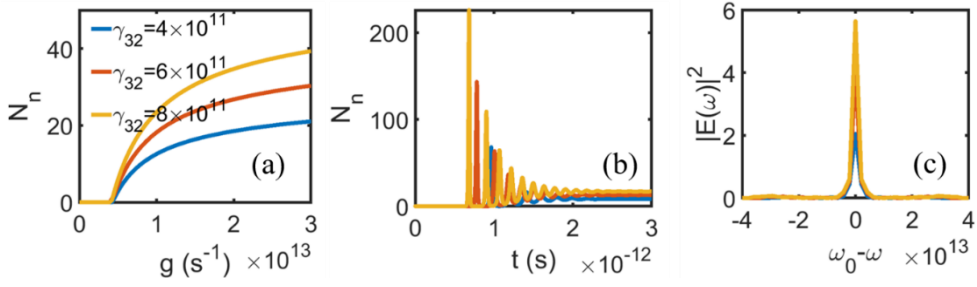

Fig. 5. (a) Number of LSP, $N_n$ as a function of pumping rate g under steady state, (b) Time evolution of $N_n$ at 8 THz pump rate, and (c) spaser spectrum at 8 THz, for different $\gamma_{32}$ values. The different values of $\gamma_{32}$ and the corresponding colours are shown in (a).

The nature of $N_n$ depends on the coherent term $\rho_{21}$, through $a_{0n}$, as seen in Eq. 8. In particular, as the imaginary value of $\rho_{21}$ increases $a_{0n}$ also increases. Fig. 6 (a) shows that the oscillation in $Im(\rho_{21})$ lasts longer for larger $\gamma_{32}$. The high amplitude oscillations in $Im(\rho_{21})$ directly transforms in $N_n$ as observed in Fig. 5 (b). So, the energy exchange between the $N_n$ and $\rho_{21}$ is the reason behind the initial oscillation in their temporal evolution but with the continuous pumping both reach to a steady state after certain time. Eq. (6) shows that the time evolutions of $\rho_{22}$ and $\rho_{11}$ depends on the coherence term $\rho_{21}$. Therefore, the time evolutions of $\rho_{22}$ and $\rho_{11}$ also show the oscillations before reaching to steady state, as displayed in Figs. 6 (b) and (c), respectively, for different values of $\gamma_{32}$ at pump rate of 8 *THz*.

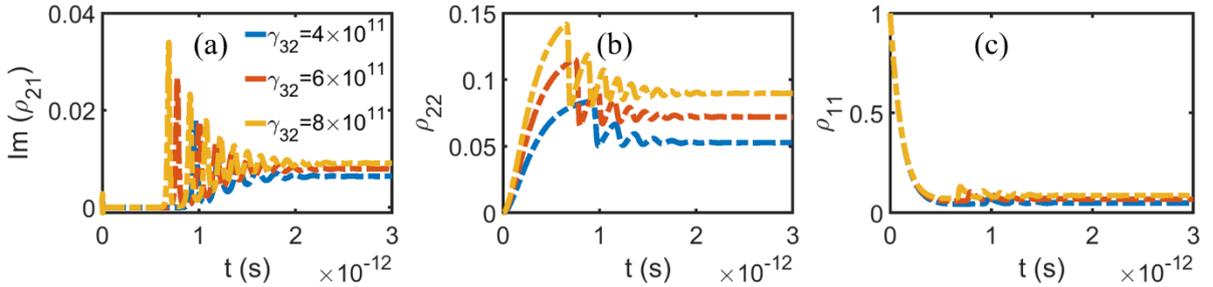

Fig. 6. Time evolution of (a) $Im(\rho_{21})$, (b) $\rho_{22}$, and (c) $\rho_{11}$ for different values of $\gamma_{32}$ at g=8 THz. The different values of $\gamma_{32}$ and the corresponding colours are shown in (a).

## IV. Conclusion

We presented a detail study of the electron-phonon interaction effects on a three-level QD-based spaser. In particular, two types of interaction potentials were discussed. The first one was linear, which resulted in time-harmonic oscillation in continuous spaser mode. These oscillations were found to depend on the Frohlich constant. This process allows for the generation of coherent phonons, which may be useful in realizing phonon nanolaser. The temporal profile of number of phonons showed decaying time harmonic oscillations, which relaxed to a steady state under continuous pumping. The number of phonons found to depend on the linear potential strength while the time period of oscillation depends on the phonon frequency. The phonon spectrum showed two resonance peaks- the peak with small amplitude appeared at the phonon resonance frequency, which indicates that the crystal lattice vibrational or phonon modes are resonating at their natural frequency. The other peak has a high amplitude and was located on the blue side of the spectrum. This peak is attributed to an electron-phonon interaction involving energy level $|2\rangle$. This interaction causes an enhanced absorption or emission of phonon energy at this frequency. The second type of potential that we discussed was quadratic potential, whose role was delineated by the decay rate $\gamma_{32}$. As the value of $\gamma_{32}$ increases, the population of level $|2\rangle$ also increases, which resulted in a large number of LSP in the spaser. The time evolution of $N_n$ shows that larger $\gamma_{32}$ increases the amplitude and time duration of the oscillations, which relaxes to a large number of $N_n$ in continuous spaser mode. A high amplitude spaser field was observed for large $\gamma_{32}$ values. In summary, this study offers more clarity on the experimental results of spaser and develops better understanding of the underlying physics.

**Acknowledgements-** Ankit Purohit is grateful to the University grant commission, India, for awarding the senior research fellowship.